\documentstyle[12pt]{article}

\newcommand{\sect}[1]{\setcounter{equation}{0}\section{#1}}

\newcommand{\app}[1]{\setcounter{section}{0}
\setcounter{equation}{0} \renewcommand{\thesection}{\Alph{section}}
\section{#1}}

\newcommand{\eq}{\begin{equation}}
\newcommand{\eqa}{\begin{eqnarray}}
\newcommand{\en}{\end{equation}}
\newcommand{\ena}{\end{eqnarray}}
\newcommand{\enn}{\nonumber \end{equation}}

\def\square{\,\lower0.9pt\vbox{\hrule \hbox{\vrule height 0.2 cm
\hskip 0.2 cm \vrule height 0.2 cm}\hrule}\,}

\def\sk{\vskip .4cm}
\def\noi{\noindent}
\def\om{\omega}
\def\al{\alpha}
\def\be{\beta}

\def\ram{{1 \over {r-r^{-1}}}}

\def\Cb{\bar{C}}

\def\epsi{\varepsilon}
\def\we{\wedge}

\def\de{\delta}

\def\part{\partial}

\def\R#1#2{ R^{#1}_{~~~#2} }

\def\Rinv#1#2{ (R^{-1})^{#1}_{~~~#2} }

\def\Rbo{{\bf R}}

\def\Rh{{\hat R}}

\def\Rhat#1#2{ \Rh^{#1}_{~~~#2} }
\def\L#1#2{ \La^{#1}_{~~~#2} }
\def\Linv#1#2{ (\La^{-1})^{#1}_{~~~#2} }

\def\Rhatinv#1#2{ (\Rh^{-1})^{#1}_{~~~#2} }

\def\La{\Lambda}

\def\cchi#1#2{\chi^{#1}_{~#2}}
\def\ome#1#2{\om_{#1}^{~#2}}
\def\RRhat#1#2#3#4#5#6#7#8{\La^{~#2~#4}_{#1~#3}|^{#5~#7}_{~#6~#8}}

\def\LL#1#2#3#4#5#6#7#8{\La^{~#2~#4}_{#1~#3}|^{#5~#7}_{~#6~#8}}

\def\Cb{{\bf C}}
\def\CC#1#2#3#4#5#6{\Cb_{~#2~#4}^{#1~#3}|_{#5}^{~#6}}

\def\C#1#2{ {\bf C}_{#1}^{~~~#2} }
\def\c#1#2{ C_{#1}^{~~~#2} }
\def\Cmat#1#2#3#4{C^{#1~~#3}_{~#2~~#4}}
\def\Cmatinv#1#2#3#4{C_{#1~~#3}^{~#2~~#4}}

\def\Dmat#1#2{D^{#1}_{~#2}}

\def\Smat#1#2{S^{~#2}_{#1}}
\def\Smatinv#1#2{(S^{-1})^{~#2}_{#1}}

\def\f#1#2{ f^{#1}_{~~#2} }

\def\T#1#2{ T^{#1}_{~~#2} }

\def\Lt#1#2{ {\tilde \Lambda}^{#1}_{~~#2} }
\def\Lbar#1#2{ {\bar \Lambda}^{#1}_{~~#2} }
\def\Ct#1#2{ {\tilde {\bf C}}_{#1}^{~~#2} }
\def\Cbar#1#2{ {\bar {\bf C}}_{#1}^{~~~#2} }

\def\M#1#2{ M_{#1}^{~#2} }
\def\qm{q^{-1}}

\def\fm#1#2{f^{-#1}_{~~~#2}}

\def\D{\Delta}

\def\Dp{\Delta^{\prime}}
\def\Ip{I^{\prime}}
\def\ep{\epsi^{\prime}}
\def\kp{k^{\prime}}
\def\lp{l^{\prime}}
\def\ip{i^{\prime}}
\def\jp{j^{\prime}}

\def\qh{q^{1\over 2}}
\def\qmh{q^{-{1\over 2}}}

\def\qm{q^{-1}}
\def\qmt{q^{-2}}

\def\n2{{{N+1} \over 2}}
\def\ap{a^{\prime}}
\def\bp{b^{\prime}}
\def\cp{c^{\prime}}
\def\Dc{{\cal D}}
\def\osqrt{{1 \over \sqrt{2}}}

\def\lie#1{\ell_{t_{#1}}}
\def\con#1{i_{t_{#1}}}

\begin{document}

\begin{titlepage}
\rightline{DFTT-70/93}
\vskip 2em
\begin{center}{\bf  DIFFERENTIAL CALCULUS ON $ISO_q(N)$,
\\ QUANTUM POINCAR\'E
ALGEBRA AND q-GRAVITY}
\\[6em]
 Leonardo Castellani
\\[2em]
{\sl Istituto Nazionale di
Fisica Nucleare, Sezione di Torino
\\and\\Dipartimento di Fisica Teorica\\
Via P. Giuria 1, 10125 Torino, Italy.}  \\
\sk

\vskip 2cm

\end{center}
\begin{abstract}
We present a general method to deform the inhomogeneous algebras
of the $B_n,C_n,D_n$ type, and find the corresponding bicovariant
differential calculus.
The method is based on a projection from $B_{n+1}, C_{n+1}, D_{n+1}$.
 For example we obtain the (bicovariant) inhomogeneous
$q$-algebra $ISO_q(N)$
as a consistent projection of the (bicovariant)  $q$-algebra $SO_q(N+2)$.
This projection works
for  particular multiparametric deformations of $SO(N+2)$, the so-called
``minimal" deformations. The case of  $ISO_q(4)$ is studied in detail:
a real form corresponding to a Lorentz signature exists only for
one of the minimal deformations, depending on one parameter
$q$. The quantum Poincar\'e Lie algebra is given explicitly: it has
10 generators (no dilatations) and contains
the {\sl classical} Lorentz algebra.
Only the commutation relations involving the
momenta depend on
$q$. Finally, we discuss a $q$-deformation of gravity based on
the ``gauging" of  this $q$-Poincar\'e algebra: the lagrangian
 generalizes
the usual Einstein-Cartan lagrangian.
\end{abstract}

\vskip 2cm

\noi DFTT-70/93

\noi December 1993
\vskip .2cm
\noi \hrule
\vskip.2cm
\hbox{\vbox{\hbox{{\small{\it e-mail addresses:}}}\hbox{}}
 \vbox{\hbox{{\small decnet=31890::castellani;}}
\hbox{{\small internet= castellani@to.infn.it }}}}

\end{titlepage}

\newpage
\setcounter{page}{1}

\sect{Introduction }
{}~~~Perturbative quantum Einstein gravity is known to be
mathematically inconsistent, since it is
plagued by ultraviolet divergences appearing at two-loop order
(the absence of one-loop divergencies was found in \cite{tHooft},
whereas two-loop divergencies were explicitly computed in
 \cite{Sagnotti}).  In supergravity the situation is only
slightly better,  the divergences
starting presumably at three loops \footnote{no explicit
 calculation like the one
of ref. \cite{Sagnotti} exists, but there is no symmetry principle
that excludes them.} . In the last fifteen years or so there have been
various proposals to overcome this difficulty,  and consistently
quantize gravity either alone or as part of a unified theory of
the fundamental interactions.  Such a unified picture is
provided by superstrings (see for a review \cite{GSW}),
where Einstein gravity arises as a low-energy effective theory,
coupled more or less realistically to gauge fields and leptons, and
regulated at the Planck scale by an infinite number of heavy particles
(the superstring massive spectrum).  How to make phenomenological
predictions from superstrings is still object of current research.
\sk
Another, more speculative,  line of thought deals with the quantization
of spacetime itself, whose smoothness under distances of the order of
the Planck length $L_P \sim 10^{-33} cm$ is really a mathematical
 assumption. Indeed if we
probe spacetime geometry with a test particle, the accuracy of the
measure depends on the Compton wavelength of the particle. For
higher accuracy we need a higher mass $m$ of the particle, and for
$m \sim 1/L_P$ the mass significantly modifies the curvature it is
supposed to measure (i.e the curvature radius becomes of the order of the
particle wavelength:  the
particle is no more a {\sl test} particle).
\sk
Thus it is not inconceivable that spacetime  has
an intrinsic cell-like structure: lattice gravity, or Regge calculus
may turn out to be something more fundamental than a
regularization procedure. Another way to discretization
is provided by non-commutative geometry:
when  spacetime coordinates do not commute the position
of a particle cannot be measured exactly. The notion
of spacetime {\sl point} loses its physical meaning, and is
to be replaced
by the notion of spacetime cell;  the question is whether
this sort of lattice
structure does indeed regularize gravity
at short distances.  References on non-commutative geometry
and its uses for regularization can be found in \cite{noncom}.
\sk
Fundamental interactions are described by field theories
with an underlying algebraic structure given by particular
Lie groups, as for ex. unitary Lie groups for the strong and electroweak
interactions and the Poincar\'e group for gravity. It is natural
to consider the so-called quantum groups \cite{qgroups1,FRT,Majid}
 (continuous
deformations of Lie groups whose geometry is non-commutative)
as the algebraic basis for generalized gauge and gravity
theories.  The bonus is that we maintain a rich algebraic structure,
more general than Lie groups, in a theory living in a
discretized space. This does not happen usually with lattice
 approaches, where one loses the symmetries of the continuum.
For a review of non-commutative differential geometry
on quantum groups see for ex. \cite{Aschieri1}.
This subject, initiated in \cite{Wor}, has been actively
 developed in recent
years: a very short list of references can be found
in \cite{Bernard}-\cite{Schupp}.
\sk
In this paper we address the problem of constructing a
non-commutative deformation of Einstein gravity. For
this we need a $q$-deformation of the Poincar\'e
Lie algebra. We obtain it in Section 4 as a special case of the
quantum  inhomogeneous $ISO_q(N)$ algebras, whose
differential calculus is presented in Section 3.  These
algebras are obtained as
projections from particular
multiparametric deformations of $SO(N+2)$,
called ``minimal" deformations. Their $R$ matrix is
{\it diagonal}, and the braiding matrix $\Rh=PR$ has
unit square. On $q$-groups with diagonal $R$-matrices see
for ex. \cite{Zupnik} and references therein.
 Deformations of Lie algebras whose
braiding matrix has unit square
were considered some time
ago by Gurevich \cite{Gurevich}.
\sk
The projective method to obtain the bicovariant differential
calculus on inhomogeneous quantum groups was introduced
in \cite{CasIGL} for $IGL_q(N)$. References on inhomogeneous
$q$-groups can also be found in \cite{inhom}.
\sk
 A general discussion
on the differential calculus on multiparametric $q$-groups is given in
Section 2.  In Section 5 we discuss the $q$-deformation
of Cartan-Maurer equations, Bianchi identities, diffeomorphisms
and propose a lagrangian for $q$-gravity, based on $ISO_q(3,1)$.
Other deformations of the Poincar\'e algebra have been considered
in recent literature \cite{qpoincare}. Although interesting
in their own right, none of these deformations corresponds
to a bicovariant differential calculus on a quantum Poincar\'e
group.

\sect{Bicovariant calculus on multiparametric quantum groups}

We recall that (multiparametric)
quantum groups are characterized by their $R$-matrix, which
 controls the noncommutativity of the quantum group
 basic elements $\T{a}{b}$ (fundamental representation):
\eq
\R{ab}{ef} \T{e}{c} \T{f}{d} = \T{b}{f} \T{a}{e} \R{ef}{cd} \label{RTT}
\en
and satisfies the quantum Yang-Baxter equation
\eq
\R{a_1b_1}{a_2b_2} \R{a_2c_1}{a_3c_2} \R{b_2c_2}{b_3c_3}=
\R{b_1c_1}{b_2c_2} \R{a_1c_2}{a_2c_3} \R{a_2b_2}{a_3b_3}, \label{QYB}
\en
a sufficient condition for the consistency of the
``RTT" relations (\ref{RTT}).  The $R$-matrix components $\R{ab}{cd}$
depend
continuously on a (in general complex)
set of  parameters $q_{ab},r$.  For $q_{ab}=q, r=q$ we
recover the uniparametric $q$-groups of ref. \cite{FRT}. Then
$q_{ab} \rightarrow 1, r \rightarrow 1$ is the classical limit for which
$\R{ab}{cd} \rightarrow \de^a_c \de^b_d$ : the
matrix entries $\T{a}{b}$ commute and
become the usual entries of the fundamental representation. The
multiparametric $R$ matrices for the $A,B,C,D$ series can be
found in \cite{Schirrmacher}  (other ref.s on multiparametric
$q$-groups are given in \cite{multiparam}). For
the $B,C,D$ case they read:

\eq
\begin{array}{ll}
\R{ab}{cd}=&\delta^a_c \delta^b_d (1-\delta^{an_2}) [r \delta^{ab}
+r^{-1} \delta^{a \bp} + (1-\de^{a\bp}) ({r\over q_{ab}} \theta (b,a)+
{q_{ba} \over r} \theta (a,b))]  \\
&+(r-r^{-1})
[\theta(a,b) \delta^b_c \de^a_d - \theta (a,c) r^{\rho_a - \rho_c}
\de^{b\ap} \de_{d\cp}]
+\de^a_{n_2} \de^b_{n_2} \de^{n_2}_c \de^{n_2}_d
\end{array}
\label{Rmp}
\en
\noi where $\theta(x)=1$ for $x > 0$
and $\theta(x)=0$ for $ x \le 0$; we define
 $n_2 \equiv \n2$ and primed indices as $\ap \equiv N+1-a$. The indices
run on $N$ values ($N$=dimension of
the fundamental representation $\T{a}{b}$),
with $N=2n+1$ for $B_n  [SO(2n+1)]$,  $N=2n$
for $C_n [Sp(2n)]$, $D_n [SO(2n)]$.
The terms with the index $n_2$ are present
only for the $B_n$ series. The
$\rho$ vector is given by:

\eq
(\rho_1,...\rho_N)=\left\{ \begin{array}{ll}
         (n- {1\over 2}, n-{3\over 2},...,
{1\over 2},0,-{1\over 2},...,-n+{1\over 2})
                   & \mbox{for $B_n$} \\
           (n,n-1,...1,-1,...,-n) & \mbox{for $C_n$} \\
           (n-1,n-2,...,1,0,0,-1,...,-n+1) & \mbox{for $D_n$}
                                             \end{array}
                                    \right.
\en
Moreover the following relations reduce the number of independent
$q_{ab}$ parameters \cite{Schirrmacher}:
\eq
q_{aa}=1,~~q_{ba}={r^2 \over q_{ab}}; \label{qab1}
\en
\eq
q_{ab}={r^2 \over q_{a\bp}}={r^2 \over q_{\ap b}}=q_{\ap\bp}~~\mbox
{\small
($a < b$)}  \label{qab2}
\en
\noi so that the $q_{ab}$ with $a < b < {N\over 2}$ give all the $q$'s.
\sk
{\sl Remark 1:}  if we denote by
$q,r$ the set of parameters $q_{ab},r$, we
have
\eq
R^{-1}_{q,r}=R_{q^{-1},r^{-1}} \label{Rprop1}
\en
\noi The inverse $R^{-1}$ is defined by
$\Rinv{ab}{cd} \R{cd}{ef}=\de^a_e \de^b_f=\R{ab}{cd}
\Rinv{cd}{ef}$.
Eq. (\ref{Rprop1}) implies
that for $|q|=|r|=1$, ${\bar R}=R^{-1}$.
\sk
{\sl Remark 2:} for $r=1$, $\Rh^2=1$
where $\Rhat{ab}{cd} \equiv \R{ba}{cd}$.
\sk
Orthogonality  conditions can be
imposed on the elements $\T{a}{b}$, consistently
with  the $RTT$ relations (\ref{RTT}):
\eqa
& &\T{a}{b} C^{bc} \T{d}{c}= C^{ad} \nonumber\\
& &\T{a}{b} C_{ac} \T{c}{d}=C_{bd} \label{Torthogonality}
\ena
\noi where the (antidiagonal) metric is :

\eq
C_{ab}=\epsilon_a r^{\rho_a} \de_{a\bp} ~\mbox{with}~\epsilon_a=
\left\{ \begin{array}{ll} +1 & \mbox{for $B_n$} ,\\
                            +1 & \mbox{for $C_n$ and $a \le n$},\\
                              -1  & \mbox{for $C_n$ and $a > n$}.
           \end{array}
                                    \right.
\label{metric}
\en
\noi and its inverse $C^{ab}$
satisfies $C^{ab} C_{bc}=\de^a_c=C_{cb} C^{ba}$.
We see
that for the orthogonal series, the matrix elements of the metric
and the inverse metric coincide,
while for the symplectic series there is a change of sign.

The consistency of (\ref{Torthogonality}) with the $RTT$ relations
is due to the identities:
\eq
C_{ab} \Rhat{bc}{de} = \Rhatinv{cf}{ad} C_{fe} \label{crc1}
\en
\eq
 \Rhat{bc}{de} C^{ea}=C^{bf} \Rhatinv{ca}{fd} \label{crc2}
\en
\noi These identities
 hold also for $\Rh \rightarrow \Rh^{-1}$.
 The co-structures of the $B,C,D$ multiparametric quantum
groups have the same form as in the uniparametric case:
the coproduct
$\D$, the counit $\epsi$ and the coinverse $\kappa$ are given by
\eqa
& & \D(\T{a}{b})=\T{a}{b} \otimes \T{b}{c}  \label{cos1} \\
& & \epsi (\T{a}{b})=\delta^a_b\\
& & \kappa(\T{a}{b})=C^{ac} \T{d}{c} C_{db}
\label{cos2}
\ena
A conjugation can be defined trivially as $T^*=T$ or via the metric as
$T^*=(\kappa(T))^t$. In the first case, compatibility with the
$RTT$ relations (\ref{RTT}) requires ${\bar R}_{q,r}=R_{q^{-1},r^{-1}}$,
i.e. $|q|=|r|=1$, and the corresponding real forms are
$SO_{q,r}(N;\Rbo)$, $SO_{q,r}(n,n;\Rbo)$ and $Sp_{q,r}(n;\Rbo)$.
 In the second case the condition on $R$ is
${\bar \R{ab}{cd}}=\R{dc}{ba}$, which
happens for $q_{ab} {\bar q}_{ab}=r^2
\in $ {\bf R}. The metric on a ``real" basis has compact signature
$(+,+,...+)$ so that the real form is  $SO_{q,r}(N;\Rbo)$.
\sk
There is also a third way to define a conjugation on the orthogonal
quantum groups $SO_{q,r}(2n,{\bf C})$, which extends to the
multiparametric case the one
proposed by
the authors of ref. \cite{Firenze1} for $SO_{q}(2n,{\bf C})$.The
conjugation is defined by:
\eq
(\T{a}{b})^*={\cal D}^a_{~c} \T{c}{d}
{\cal D}^d_{~b} \label{Tconjugation}
\en
\noi ${\cal D}$ being the matrix that
exchanges the index n with the index n+1.
This conjugation is compatible with the coproduct: $\D (T^*)=(\D T)^*$;
for $|r|=1$ it is also  compatible with the
orthogonality relations (\ref{Torthogonality}) (due to
${\bar C}=C^T$ and also $\Dc C \Dc = C$) and with the antipode:
$\kappa(\kappa(T^*)^*)=T$.  Compatibility with the $RTT$ relations
is easily seen to require
\eq
({\bar R})_{n \leftrightarrow n+1}=R^{-1}, \label{Rprop2}
\en
\noi which implies

i) $|q_{ab}|=|r|=1$
for a and b both different from n or n+1;

ii) $q_{ab}/r \in {\bf R}$
when at least one of the indices a,b is equal
to n or n+1.
\sk

Since later we consider the case $r=1$ and
$(R)_{n \leftrightarrow n+1}=R$ (and
therefore ${\bar R}=R^{-1}$ because of
(\ref{Rprop2})), the conditions
on the parameters will be:
\eqa
& &  |q_{ab}|=1 \mbox{~~~for a and b both
different from n or n+1} \nonumber\\
& &  q_{ab}=1 \mbox{~~~for a or b equal to n or n+1}
\ena
\sk
\noi This last conjugation leads to the
real form $SO_{q,r}(n+1,n-1;\Rbo)$, and
will in fact be the one we need in order to obtain $ISO_q(3,1;\Rbo)$, as
we discuss in Section 4.
\sk
A bicovariant differential calculus \cite{Wor} on the
multiparametric $q$-groups can be
constructed in
terms of the corresponding $R$ matrix , in much the same
way as for uniparametric $q$-groups (for
which we refer to \cite{Jurco,Zumino,Aschieri1}).
Here we concentrate on  $SO_{q,r}(N+2)$, but
everything holds also for $Sp_{q,r}(N+2)$.
For later convenience we adopt upper case indices for the
fundamental representation
of $SO_{q,r}(N+2)$ and lower case indices for the fundamental
 representation of $SO_{q,r}(N)$.

The basic object is the braiding matrix
\eq
\RRhat{A_1}{A_2}{D_1}{D_2}{C_1}{C_2}{B_1}{B_2}
\equiv  d^{F_2} d^{-1}_{C_2} \R{F_2B_1}{C_2G_1} \Rinv{C_1G_1}{E_1A_1}
    \Rinv{A_2E_1}{G_2D_1} \R{G_2D_2}{B_2F_2} \label{Lambda}
\en
which is used in the definition of the exterior product of
quantum left-invariant one forms $\ome{A}{B}$:
\eq
\ome{A_1}{A_2} \we \ome{D_1}{D_2}
\equiv \ome{A_1}{A_2} \otimes \ome{D_1}{D_2}
- \RRhat{A_1}{A_2}{D_1}{D_2}{C_1}{C_2}{B_1}{B_2}
\ome{C_1}{C_2} \otimes \ome{B_1}{B_2} \label{exteriorproduct}
\en
\noi and in the $q$-commutations of the quantum Lie
algebra generators $\cchi{A}{B}$:
\eq
\cchi{D_1}{D_2} \cchi{C_1}{C_2} - \RRhat{E_1}{E_2}{F_1}{F_2}
{D_1}{D_2}{C_1}{C_2} ~\cchi{E_1}{E_2} \cchi{F_1}{F_2} =
\CC{D_1}{D_2}{C_1}{C_2}{A_1}{A_2} \cchi{A_1}{A_2}
\label{qLie}
\en
\noi where the structure constants are explicitly given by:
\eq
\CC{A_1}{A_2}{B_1}{B_2}{C_1}{C_2} =\ram [- \de^{B_1}_{B_2}
\de^{A_1}_{C_1}
\de^{C_2}_{A_2} + \RRhat{B}{B}{C_1}{C_2}{A_1}{A_2}{B_1}{B_2}]. \label{CC}
\en
The $d^A$ vector in (\ref{Lambda})
is defined via the diagonal matrix  $\Dmat{A}{B}$ as
$d^A= \Dmat{A}{A}$ (no sum on $A$), with $D=CC^t$,or
\eq
\Dmat{A}{B}=C^{AC} C_{BC} \label{Dmatrix}
\en
A graphical representation of the braiding matrix (\ref{Lambda}) is
given in Appendix A.
\sk
{\sl Remark 3:} for $r=1$ we have $\Lambda^2=1$. This is due to $\Rh^2=1$
and $\Dmat{A}{B}=\de^A_B$.
\sk
The braiding matrix $\La$  and
the structure constants $\Cb$ defined in
(\ref{CC}) satisfy
the conditions

\eqa
& & \C{ri}{n} \C{nj}{s}-\L{kl}{ij} \C{rk}{n} \C{nl}{s} =
\C{ij}{k} \C{rk}{s}
{}~~\mbox{({\sl q}-Jacobi identities)} \label{bic1}\\
& & \L{nm}{ij} \L{ik}{rp} \L{js}{kq}=\L{nk}{ri} \L{ms}{kj}
\L{ij}{pq}~~~~~~~~~\mbox{(Yang--Baxter)} \label{bic2}\\
& & \C{mn}{i} \L{ml}{rj} \L{ns}{lk} + \L{il}{rj} \C{lk}{s} =
\L{pq}{jk} \L{is}{lq} \C{rp}{l} + \C{jk}{m} \L{is}{rm}
\label{bic3}\\
& & \C{rk}{m} \L{ns}{ml} = \L{ij}{kl} \L{nm}{ri} \C{mj}{s}
\label{bic4}
\ena

\noi where the index pairs ${}_A^{~B}$ and ${}^A_{~B}$
have been replaced by the indices ${}^i$ and ${}_i$ respectively.
These are the so-called ``bicovariance conditions", see ref.s
\cite{Wor,Bernard,Aschieri1},
necessary for the existence of a consistent bicovariant
differential  calculus, as we discuss further in Appendix B.
\sk
A metric can be defined in the adjoint representation of the
$B_n,C_n,D_n$ $q$-groups
as follows:
\eq
C_{ij} \equiv \Cmat{c_1}{c_2}{b_1}{b_2}=C^{c_1f}
\Rinv{b_1e}{fc_2} C_{b_2e}
\label{metricadjoint}
\en
\eq
C^{ij} \equiv \Cmatinv{a_1}{a_2}{c_1}{c_2}=C_{a_1e}
\R{ea_2}{c_1f} C^{c_2f}
\label{metricadjointinv}
\en
\noi  and satisfies the relations:
\eq
C_{ij} C^{jk}=\de_i^k=C^{kj} C_{ji}
\en
\noi and
\eq
C^{ik} \L{sl}{kr}=\Linv{is}{rj} C^{jl} \label{CRC1}
\en
\eq
\L{rj}{is} C_{jl}=C_{ik} \Linv{kr}{sl} \label{CRC2}
\en
\noi i.e. the analogue of eq.s (\ref{crc1})-(\ref{crc2}). These
relations allow
to define consistent orthogonality relations for the $q$-group matrix
elements in the adjoint representation (see Appendix  B).
\sk
{\sl Remark 4}: when $r=1$ ($ \Rightarrow \Rh^2=1, \Lambda^2=1$),
 the following useful identities hold:
\eq
D^a_{~b} \equiv C^{ac} C_{bc}=\de^a_b,~~(D^{-1})^a_{~b} \equiv C^{ca}
C_{cb}=\de^a_b
\en
\eq
D^i_{~j }\equiv C^{ik} C_{jk}=
\de^i_j,~~(D^{-1})^i_{~j} \equiv C^{ki} C_{kj}=\de^i_j
\en
\eq
\Rhat{ab}{cd} \Rhat{ce}{af}=\de^b_f \de^e_d=\Rhat{ba}{dc} \Rhat{ec}{fa}
\en
\eq
\L{ri}{sl} \L{sk}{rj}=\de^i_j \de^k_l =\L{ir}{ls}
\L{ks}{jr}\label{Ltilde}
\en
\sk
The first two $*$-conjugations (the  ``usual ones")
on the $T$'s we have discussed
earlier in this Section
can be extended to the dual space spanned by the
$q$-Lie algebra generators $\chi$  as in the uniparametric case.
The consistent extension of the third conjugation to
the $\chi$ space is treated in Appendix C,
for the case of minimal deformations ($r=1$) of $SO(2n)$. We find that
\eq
(\cchi{a}{b})^*=-{\cal D}^a_{~c} \cchi{c}{d} {\cal D}^d_{~b}
\label{chiconjugation}
\en
\noi is compatible with the bicovariant differential calculus if
the $\Lambda$ and $\Cb$ tensors are invariant under the exchange of the
indices n and n+1, and if the following relation holds:
\eq
{\bar \C{ij}{k}}=-\C{ji}{k} \label{conditiononC}
\en

\sect{Inhomogeneous quantum groups and their differential calculus}
In this Section we present a general method of quantizing inhomogeneous
groups whose homogeneous subgroup belongs to the $BCD$ series.
In particular we concentrate on the $q$-deformations of the $ISO(N)$
groups, as these are the groups relevant for the construction
of $q$-gravity theories.

The idea is to project
$SO_q(N+2)$ and its differential calculus on $ISO_q(N)$,
much as we did for $IGL_q(N)$ in ref. \cite{CasIGL}, where we projected
from $GL_q(N+1)$.

For this we have to consider the multiparametric deformations of the
orthogonal groups  $SO_{q,r}(N+2)$ with $r=1$ (minimal deformations).
 Only for $r=1$ we can obtain a consistent projection on
 $ISO_q(N)$.
\sk
We know that the $\R{ab}{cd}$ matrix of $SO_{q,r}(N)$ is contained in the
$\R{AB}{CD}$ matrix of
$SO_{q,r}(N+2)$: more precisely it is obtained from
the ``mother " $R$ matrix by restricting its indices to the values
{\small A,B,..=2,3,...N-1}.  We therefore split the capital indices as
{\small A}=$(\circ, a, \bullet)$. Then
the   $R$ matrix of $SO_{q,r}(N+2)$ can be
rewritten in terms of $SO_{q,r}(N)$ quantities:
\eq
\R{AB}{CD}=\left(  \begin{array}{cccccccccc}
   {}&\circ\circ&\circ\bullet&\bullet
          \circ&\bullet\bullet&\circ d&\bullet d
      &c \circ&c\bullet&cd\\
   \circ\circ&r&0&0&0&0&0&0&0&0\\
   \circ\bullet&0&r^{-1}&0&0&0&0&0&0&0\\
   \bullet\circ&0&f(r)&r^{-1}&0&0&0&0&0&-C_{cd} \lambda\\
\bullet\bullet&0&0&0&r&0&0&0&0&0\\
\circ b&0&0&0&0&{r\over q_{\circ b}} \de^b_d&0&0&0&0\\
\bullet b&0&0&0&0&0&{q_{\circ b} \over r} \de^b_d&0&\lambda\de^b_c&0\\
a\circ&0&0&0&0&\lambda\de^a_d&0&{q_{\circ a} \over r} \de^a_c&0&0\\
a\bullet&0&0&0&0&0&0&0&{r\over q_{\circ a}} \de^a_c&0\\
ab&0&-C^{ab} \lambda&0&0&0&0&0&0&\R{ab}{cd}\\
\end{array} \right)
\en
\noi where $C_{ab}$ is the $SO_{q,r}(N)$
metric,  $\lambda \equiv r-r^{-1}$
and $f(r) \equiv \lambda (1-r^{-N})$ for the
orthogonal series $B,D$  ($f(r) \equiv
\lambda (1-r^{-N-2})$ for the symplectic
series).

 It is not difficult to reexpress the
$\Lambda$ and $\Cb$ tensors in our index convention. Less trivial
is to find a subset of these components, containing the
$\Lambda$ and $\Cb$ tensors of $SO_{q,r}(N)$,  that satisfies the
bicovariance conditions (\ref{bic1}) - (\ref{bic4}).
This subset in fact exists for $r=1$ and is given by:
\eqa
& &\LL{a_1}{a_2}{d_1}{d_2}{c_1}{c_2}{b_1}{b_2}=
\R{f_2b_1}{c_2g_1} \Rinv{c_1g_1}{e_1a_1}
    \Rinv{a_2e_1}{g_2d_1} \R{g_2d_2}{b_2f_2} \label{L1}\\
& &\LL{a_1}{\circ}{d_1}{d_2}{c_1}{c_2}{b_1}{\circ}=
{q_{\circ d_1 }\over q_{\circ d_2}}
\R{d_2b_1}{c_2g_1} \Rinv{c_1g_1}{d_1a_1}
     \label{L2}\\
& &\LL{a_1}{a_2}{d_1}{\circ}{c_1}
{\circ}{b_1}{b_2}={q_{\circ b_2} \over q_{\circ b_1}}
 \Rinv{c_1b_1}{e_1a_1} \Rinv{a_2e_1}{b_2d_1}  \label{L3}\\
& &\LL{a_1}{\circ}{d_1}{\circ}{c_1}{\circ}{b_1}{\circ}=
   {q_{\circ d_1 }\over q_{\circ b_1}}
\Rinv{c_1b_1}{d_1a_1}     \label{L4}\\
& &\LL{\bullet}{a_2}{d_1}{d_2}{c_1}{c_2}{\bullet}{b_2}=
{q_{\circ c_1 }\over q_{\circ c_2}}
\Rinv{a_2c_1}{g_2d_1} \R{g_2d_2}{b_2c_2}
     \label{L5}\\
& &\LL{a_1}{a_2}{\bullet}{d_2}{\bullet}{c_2}{b_1}{b_2}=
{q_{\circ a_2} \over q_{\circ a_1}}
 \R{f_2b_1}{c_2a_1} \R{a_2d_2}{b_2f_2}  \label{L6}\\
& &\LL{\bullet}{a_2}{\bullet}{d_2}{\bullet}{c_2}{\bullet}{b_2}=
   {q_{\circ a_2 }\over q_{\circ c_2}}\R{a_2d_2}{b_2c_2}     \label{L7}\\
& &\LL{\bullet}{a_2}{d_1}{\circ}{c_1}{\circ}{\bullet}{b_2}=
   q_{\circ c_1 } q_{\circ b_2} \Rinv{a_2c_1}{b_2d_1}     \label{L8}\\
& &\LL{a_1}{\circ}{\bullet}{d_2}{\bullet}{c_2}{b_1}{\circ}=
   (q_{\circ a_1 } q_{\circ d_2} )^{-1}
\Rinv{d_2b_1}{c_2a_1}     \label{L9}
\ena
\eqa
& &\CC{c_1}{c_2}{b_1}{b_2}{d_1}{d_2}=
{\rm structure~constants~of~}SO_{q,r=1}(N) \label{C1}\\
& &\CC{c_1}{\circ}{b_1}{b_2}{d_1}{\circ}=\lim_{r \rightarrow 1}
{1 \over {r-r^{-1}}} [-\de^{b_1}_{b_2} \de^{c_1}_{d_1} +
 {q_{\circ b_2} \over q_{\circ b_1}}\Rinv{c_1b_1}{e_1a}
\Rinv{ae_1}{b_2d_1} ] \label{C2}\\
& &\CC{c_1}{c_2}{b_1}{\circ}{d_1}{\circ}=
\R{g_2b_1}{c_2g_1}
\Rinv{c_1g_1}{e_1a}\Rinv{ae_1}{g_2d_1} \label{C3}\\
& &\CC{c_1}{c_2}{b_1}{\circ}{\bullet}{d_2}=q_{\circ d_2}^{-1} C^{ae_1}
\R{d_2b_1}{c_2g_1}
\Rinv{c_1g_1}{e_1a} \label{C4}\\
& &\CC{c_1}{c_2}{\bullet}{b_2}{\bullet}{d_2}=
{q_{\circ c_1} \over q_{\circ c_2}}
\R{c_1d_2}{b_2c_2}  \label{C5}\\
& &\CC{c_1}{c_2}{\bullet}{b_2}{d_1}{\circ}=
-{q_{\circ c_1} \over {q_{\circ c_2} q_{\circ d_2}}}
C_{b_2 c_2} \de^{c_1}_{d_1}  \label{C6}\\
& &\CC{\bullet}{c_2}{b_1}{b_2}{\bullet}{d_2}=\lim_{r \rightarrow 1}
{1 \over {r-r^{-1}}} [-\de^{b_1}_{b_2} \de^{d_2}_{c_2} +
\R{f_2b_1}{c_2a}  \R{ad_2}{b_2f_2} ] \label{C7}
\ena

This is the key result of this
Section, and enables the consistent projection
on the $ISO_q(N)$ algebra by setting:
\eq
\cchi{\circ}{b}=\cchi{a}{\bullet}=
\cchi{\circ}{\circ}=\cchi{\bullet}{\bullet}=
\cchi{\circ}{\bullet}=\cchi{\bullet}{\circ}=0 \label{proj}
\en
The $ISO_q(N)$ Lie algebra is given
explicitly in Table 1. The reason we call
this the $ISO_q(N)$ Lie algebra will be explained below.
\sk
We prove now that the components (\ref{L1})-(\ref{C7}) indeed satisfy
the bicovariance conditions (\ref{bic1})- (\ref{bic4}). We label by the
letter $H$ the subset of indices
present in eq.s (\ref{L1}) - (\ref{C7}), i.e.
${}_H={}^a_{~b}, {}^a_{~\circ}, {}^{\bullet}_{~b}$
(${}^H={}_a^{~b},{}_a^{~\circ},
{}_{\bullet}^{~b}$), and by the letter $K$
all the other composite indices. We have to prove that, setting equal to
 $H$ all free
indices in (\ref{bic1}) - (\ref{bic4}),  only
$H$ indices enter in the index sums
(and therefore the $H$-tensors of
(\ref{L1})-(\ref{C7}) satisfy {\sl by themselves} the
bicovariance conditions).  This is true i) for
the quantum Yang-Baxter eqs.
(\ref{bic2}) since the tensor $P\Lambda$ is diagonal for $r=1$, so that
\eq
\L{HH}{HK}=\L{HH}{KH}=\L{HH}{KK}=\L{HK}{HH}=\L{KH}{HH}=\L{KK}{HH}=0;
\label{LHK}
\en
ii) for the $q$-Jacobi eqs. (\ref{bic1})
because $\C{HH}{K}$ can be different
from zero only when ${}^K={}_{\bullet}^{~\circ}$, and
$\C{HK}{H}=\C{KH}{H}=0$
when ${}_K ={}^{\bullet}_{~\circ}$;
iii) for the last two bicovariant conditions
(\ref{bic3}) - (\ref{bic4}) again because of (\ref{LHK}).  \square
\sk
Thus far we have shown that
there is a subset of $\cchi{A}{B}$ (the generators
of the $q$-Lie algebra of
$SO_{q,r=1}(N+2)$ )  closing on the $q$-algebra
of Table 1, namely $\cchi{a}{b}, \cchi{a}{\circ} $
and  $\cchi{\bullet}{b}$.
This algebra is bicovariant, in the sense that the corresponding
$\Lambda$ and $\Cb$ tensors satisfy (\ref{bic1})-(\ref{bic4}) . It would
seem that the number of  momenta is twice what we need, since there
are two kinds of  ``momentum" generators,
$\cchi{a}{\circ} $and  $\cchi{\bullet}{b}$.
However by examining in some detail the $q$-algebra we can
conclude that only $N$ combinations of these momenta do survive,
and if we rewrite the algebra of Table 1 in terms of these combinations
we precisely obtain a deformation of $ISO(N)$. Let us prove this.
\sk
Consider the structure constants $\CC{c_1}{\circ}{b_1}{b_2}{d_1}{\circ}$.
It is not difficult to see from (\ref{C2})  that for $c_1=b_2, b_1=b_2'$
these constants are vanishing for any value of $r$ (use the explicit
expression (\ref{Rmp})), and thus in particular for $r=1$. On the other
hand the structure constants $\CC{b_1}{b_2}{c_1}{\circ}{d_1}{\circ}$
and $\CC{b_1}{b_2}{c_1}{\circ}{\bullet}{d_2}$ are {\sl not} vanishing
for the same values of $c_1,b_1,b_2$, but:
\eqa
& &\CC{b_1}{b_2}{c_1}{\circ}{d_1}{\circ}=\de^{c_1}_{d_1}\\
& &\CC{b_1}{b_2}{c_1}{\circ}{\bullet}{d_2}=C^{c_1b_1} q_{\circ d_2}^{-1}
       \de^{d_2}_{b_2}
\ena
Thus we have the two commutations:
\eq
\cchi{c_1}{\circ}
\cchi{b_1}{b_2}- \LL{e_1}{e_2}{f_1}{\circ}{c_1}{\circ}{b_1}{b_2}
\cchi{e_1}{e_2} \cchi{f_1}{\circ}=0
\en
\eq
\cchi{b_1}{b_2}\cchi{c_1}{\circ} -
\LL{e_1}{\circ}{f_1}{f_2}{b_1}{b_2}{c_1}{\circ}
\cchi{e_1}{\circ} \cchi{f_1}{f_2}=
\cchi{b_1}{\circ}+q_{\circ b_1'}^{-1} \cchi{\bullet}
{b_1'}
\en
\noi Next we remark that for
$\Lambda^2=I$ as is the case for $r=1$ the two
left-hand sides of the above equations
are equal up to a minus sign, so that
finally we have:
\eq
\cchi{b_1}{\circ}+q_{\circ b_1'}^{-1}
\cchi{\bullet}{b_1'}=0 \label{constraint}
\en
These $N$ equations reduce the number of independent momenta to $N$.
We can easily rewrite the
algebra in Table 1 in terms of the redefined momenta:
\eq
 \chi^a \equiv q_{\circ a'}^{1\over 2} \cchi{a}{\circ}-
q_{\circ a'}^{-{1\over 2}} \cchi{\bullet}{a'}
\en
\noi and we have done so in Table 2.

This was possible because the $q$-commutator of  $\cchi{a}{\circ}$ with
a given generator $\chi$  is the same
as the $q$-commutator of $-q^{-1}_{\circ a'}
\cchi{\bullet}{a'}$ with $\chi$,
 because the constraint (\ref{constraint}) is consistent with
the $q$-Lie algebra of Table 1. Another way to see it is to remark
that the algebra of Table 1
satisfies the $q$-Jacobi
identities (\ref{bic1}).  Then we have an explicit
matrix representation of the
$q$-generators $\chi$: the adjoint representation
$(\chi_i)^j_{~k} \equiv \C{ki}{j}$. Eq. (\ref{constraint}) means
that the generators $q_{\circ a'}^{1\over 2} \cchi{a}{\circ}$ and
$-q_{\circ a'}^{-{1\over 2}}
\cchi{\bullet}{a'}$ have the {\sl same} matrix
representative, and hence the same commutations with the other
generators.
\sk
The $q$-Lie algebra of Table 2 satisfies the bicovariant conditions
(\ref{bic1})-(\ref{bic4}). As discussed in Appendix C, these
define a (bicovariant) differential calculus on the quantum
group generated by the elements $\M{i}{j}$ (the $q$-group
elements in the adjoint representation). This $q$-group we
call the quantum $ISO_q(N)$ group.
\sk
Note however that we do not have an invertible adjoint metric any more
(but only a submetric $C_{ij}$
with $i,j$ along the $SO_q(N)$ directions).
Then the  existence of the antipode of $\M{i}{j}$ is not ensured via
eq. (\ref{Mantipode}). What we have
called ``the quantum $ISO_q(N)$ group"
is a Hopf algebra only if we manage to find an inverse $(M^{-1})_i^{~j}$.
Otherwise we have a bialgebra.
\sk
However (see the discussion at the end of Appendix B)
we know that we can find an antipode for the dual algebra
generated by $\chi_i,\f{i}{j}, \fm{i}{j}$,  without
reference to an adjoint metric.
For the argument, eqs. (\ref{Ltilde})
were crucial: do they hold also in the ``projected" case ?
The answer is yes, due to the matrix $P\Lambda$ being diagonal
for $r=1$. The algebra generated by the $\chi_i, \f{i}{j}, \fm{i}{j}$
is a {\sl bona fide} Hopf algebra, which we call the quantum $ISO_q(N)$
bicovariant algebra  (we reserve the name of $q$-Lie algebra
to the one generated only by the $\chi_i$).
\sk
Finally, we come to the $*$-conjugation on the generator space
induced by the rule (\ref{chiconjugation}), for $ISO_q(2n)$.
 Recall that this
rule is consistent when
 the  $\Lambda$ and
$\Cb$ tensors are $n \leftrightarrow n+1$ invariant and condition
 (\ref{conditiononC}) is satisfied. The
question is whether the projected $*$-conjugation is still compatible
with the projected differential calculus. The answer is yes:
 the $\Lambda$ and $\Cb$ tensors corresponding to
 the algebra in Table 2, satisfying the bicovariance conditions
(\ref{bic1})-(\ref{bic4}), are still
invariant under the exchange of the (fundamental) indices
n and n+1. If the structure constants {\bf C}
satisfy (\ref{conditiononC}), the result of
Appendix C holds also
for $ISO_q(2n)$.
\sk
Then we have the (Hopf algebra) conjugation:
\eq
(\cchi{a}{b})^*=-{\cal D}^a_{~c} \cchi{c}{d} {\cal D}^d_{~b}
\label{Ichiconjugation1}
\en
\eq
(\chi^a)^*=-\Dc^a_{~b} \chi^b \label{Ichiconjugation2}
\en
\noi whose consistency can be checked explicitly in the example
 of the next Section.

\sect{$ISO_q(3,1)$ and the quantum Poincar\'e Lie algebra}

We come now to applying the preceding formalism to the
case of $ISO(4)$. We know from the discussion of the previous
Section that  a real form exists corresponding to a $(3,1)$ signature.
Let us consider the ``mother" $R$-matrix of $SO_{q,r=1}(6)$.
According to (\ref{qab1}) and (\ref{qab2}) there are three independent
deformation parameters, i.e. $q_{\circ 1}$,  $q_{\circ 2}$ and $q_{12}$
 (in the index convention $a=\circ,1,2,3,4,\bullet$).
It is not difficult to see that this $R$ matrix
has the  $2 \leftrightarrow 3$ symmetry
only if  $q_{\circ 2}=1$ and $q_{12}=1$.
Therefore we are left with
the only parameter $q_{\circ 1} \equiv q$. Note that $q_{12}$ is the
deformation parameter of the Lorentz subalgebra, and $q_{12}=1$ means
that this subalgebra is {\it classical}.
\sk
Consider now the $ISO_q(4)$ algebra one deduces by specializing
$N=4$, $q_{12}= q_{\circ 2}=1, q_{\circ 1} \equiv q$ in Table 2.
The result is given in Table 3, in the basis
$\chi_a \equiv C_{ab} \chi^b$,
$\chi_{ab}\equiv {1\over 2} [C_{ac} \cchi{c}{b}-C_{bc} \cchi{c}{a}]$,
where the metric $C_{ab}$ is the classical antidiagonal metric
$C_{14}=C_{23}=C_{32}=C_{41}=1$ (otherwise $0$). The
invariance under the index exchange $2 \leftrightarrow 3$ is explicit,
and the condition (\ref{conditiononC}) is easily seen to hold.
\sk
Then we can define a  consistent  $*$-conjugation
on the $\chi$, according
to (\ref{Ichiconjugation1}), (\ref {Ichiconjugation2}):
\eqa
& &\chi_{\al \be}^*=-\chi_{\al \be}
\mbox{~~~($\al,\be \not=2,3$)} \nonumber\\
& &\chi_{2 \be}^*=-\chi_{3\be} \nonumber \\
& &\chi_{3\be}^*=-\chi_{2\be} \nonumber \\
& &\chi_{23}^*= \chi_{23}
\ena
\eq
(\chi_1)^*=-\chi_1,~(\chi_2)^*=-\chi_3,~(\chi_3)^*=-\chi_2,~
(\chi_4)^*=-\chi_4.
\en
\noi whose compatibility with the commutations of
Table 3 can be directly verified.
\sk
This conjugation allows the definition of ``antihermitian" quantum
generators $\xi$:
\eqa
& &\xi_{\al\be}=\chi_{\al\be} \nonumber\\
& &\xi_{2\be}=\osqrt (\chi_{2\be}+\chi_{3\be}) \nonumber \\
& &\xi_{3\be}=\osqrt  i (\chi_{2\be}- \chi_{3\be}) \nonumber \\
& &\xi_{23}=-i \chi_{23}
\ena
\eqa
& &\xi_{\al}=\chi_{\al}  \nonumber\\
& &\xi_{2}=\osqrt (\chi_{2}+\chi_{3}) \nonumber \\
& &\xi_{3}=\osqrt  i (\chi_{2}- \chi_{3})
\ena
On this basis the metric in Table 3 becomes
\eq
C_{ab}=\left(  \begin{array}{cccc}
     0&0&0&1\\
     0&1&0&0\\
     0&0&1&0\\
     1&0&0&0\\
      \end{array} \right)  \label{lorentzmetric}
\en
\noi with the desired signature $(+,+,+,-)$.

\sect{Cartan-Maurer equations, $q$-diffeomorphisms
and $q$-gravity}

In this Section we discuss the $q$-generalization of
Poincar\'e gravity based on the deformed Poincar\'e
algebra of Table 3. As in the classical case we start
by defining the curvatures. To do so, we first need the
deformed Cartan-Maurer equations \cite{Wor,Aschieri1}
\eq
d \om^i + \c{jk}{i} \om^j \we \om^k=0
\en
\noi where the $\om$ are the left-invariant one-forms
discussed in Appendix B. The $C$ structure constants
appearing in the Cartan-Maurer equations are in general related
to the $\Cb$ constants of the $q$-Lie algebra \cite{Aschieri1}:
\eq
\C{jk}{i}=\c{jk}{i}-\L{rs}{jk} \c{rs}{i}
\en
\noi In the particular case $\Lambda^2=I$ it is not difficult to
see that in fact
$C= {1 \over 2} \Cb$, which is a worthwhile simplification.
\sk
The procedure we have advocated in ref.s \cite{Cas1} for the
``gauging" of quantum groups essentially retraces the steps
of the group-geometric method for the
gauging of usual Lie groups, described
for instance in ref.s \cite{Cas2}.

We consider one-forms $\om^i$ which are not left-invariant any
more, so that the Cartan-Maurer equations are replaced by:
\eq
R^i=d \om^i + \c{jk}{i} \om^j \we \om^k \label{curvature}
\en
\noi where the curvatures $R^i$ are now non-vanishing, and satisfy
the $q$-Bianchi identities:
\eq
dR^i-\c{jk}{i} R^j \we \om^k + \c{jk}{i} \om^j \we R^k=0 \label{bianchi}
\en
\noi due to the Jacobi identities on the structure constants $C$
\cite{Aschieri1}. As in the classical case we can write the
 $q$-Bianchi identities as $\nabla R^i=0$, which define the
covariant derivative $\nabla$.

Eq. (\ref{curvature})
can be taken as the definition of the curvature $R^i$.  We apply
it to the $q$-Poincar\'e algebra of Table 3: the one-forms are
$\om^i \equiv V^a, \om^{ab}$ and the corresponding
curvatures read (we omit wedge symbols):
\eqa
& &R^1=dV^1 + \qmh  \om^{12}  V_2 + \qmh \om^{13} V_3+ \om^{14} V_4
\nonumber \\
& &R^2=dV^2 - \qmh  \om^{12}  V_1 + \om^{23} V_3+  \qh \om^{24} V_4
\nonumber \\
& &R^3=dV^3 - \qmh  \om^{13}  V_1 - \om^{23} V_2+ \qh \om^{34} V_4
\nonumber \\
& &R^4=dV^4-  \om^{14}  V_1- \qh \om^{24} V_2-\qh \om^{34} V_3
\ena
\eq
R^{ab}=d \om^{ab} +  C_{cd}\om^{ac}  \om^{db} \label{Lorentzcurv}
\en

\noi where  $V_a \equiv  C_{ab} V^b$,  $C_{ab}$
is given in (\ref{lorentzmetric}). We have rescaled
$\om^{ab}$ by a factor ${1\over 2}$ to obtain standard
normalizations.  $R^{ab}$ is the
$q$-Lorentz curvature, coinciding with the classical
one (as a function of
$\om^{ab}$), and $R^a$ is the $q$-deformed torsion.
\sk
{}From the definition (\ref{wedge}) of the exterior product we see that
for $\Lambda^2=I$ the one-forms $\om^i$  $q$-commute as:
\eq
\om^i \om^j =-\L{ij}{kl} \om^k \om^l
\en
Inserting the $\Lambda$ tensor corresponding to Table 3
we find:
\eqa
& &V^a \om^{12}=-\qm \om^{12} V^a \nonumber\\
& &V^a \om^{13}=-\qm \om^{13} V^a \nonumber\\
& &V^a \om^{14}=-\om^{14} V^a \nonumber\\
& &V^a \om^{23}=-\om^{23} V^a \nonumber\\
& &V^a \om^{24}=-q \om^{24} V^a\nonumber\\
& &V^a \om^{34}=-q \om^{34} V^a
\ena
\eqa
& &V^2 V^1 = - \qm V^1 V^2 \nonumber\\
& &V^3 V^1 = - \qm V^1 V^3 \nonumber\\
& &V^4 V^1 = - \qmt V^1 V^4 \nonumber\\
& &V^3 V^2 = - V^2 V^3 \nonumber\\
& &V^4 V^2 = - \qm V^2 V^4 \nonumber\\
& &V^4 V^3 = - \qm V^3 V^4
\ena
\noi and usual anticommutations between the $\om^{ab}$ (components
of the Lorentz spin connection). The exterior product of two identical
one-forms vanishes (this is not
true in general when $\Lambda^2 \not= I$).
\sk
We are now ready to write the lagrangian for the
$q$-gravity theory based on $ISO_q(3,1)$. The
lagrangian looks identical to the classical one, i.e.:
\eq
{\cal L}=R^{ab} V^c V^d \epsi_{abcd} \label{lagrangian}
\en
The Lorentz curvature $R^{ab}$, although defined as in the classical
case, has non-trivial commutations with the $q$-vielbein:
\eqa
& &V^a R^{12}=\qm R^{12} V^a \nonumber\\
& &V^a R^{13}=\qm R^{13} V^a \nonumber\\
& &V^a R^{14}=R^{14} V^a \nonumber\\
& &V^a R^{23}= R^{23} V^a \nonumber\\
& &V^a R^{24}= q R^{24} V^a \nonumber\\
& &V^a R^{34}=q  R^{34} V^a
\ena
\noi deducible from the definition
(\ref{Lorentzcurv}). As in ref. \cite{Cas1,Aschieri1},
we make the assumption that the commutations of $d\om^i$ with
the one-forms $\om^l$ are the same as those of $\c{jk}{i} \om^j \om^k$
with $\om^l$, i.e. the same as those valid for $R^i=0$.
For the definition of $\epsi_{abcd}$ in (\ref{lagrangian})
see below.
\sk
We discuss now the notion of $q$-diffeomorphisms. It is known
that there is a consistent $q$-generalization of the Lie derivative
(see ref.s \cite{Aschieri1,Aschieri2,Schupp} ) which can be
expressed as in the classical
case as:
\eq
\lie{i} = \con{i} d + d \con{i}
\en
\noi where $\con{i}$ is the $q$-contraction operator defined
in ref.s \cite{Aschieri1,Aschieri2}, with the following properties:
\eqa
& &i)~~~i_V (a)=0, \mbox{~~$a \in A$,
$V$ generic tangent vector} \nonumber\\
& &ii)~~\con{i} \om^j=\de^j_i I \nonumber\\
& &iii)~\con{i} (\theta \we \om^k)=
\con{r} (\theta) \om^l \L{rk}{li}+(-1)^p~ \theta ~\de^k_i
\mbox{~~$\theta$ generic $p$-form}
\nonumber\\
& &iv)~~i_V(a \theta + \theta')=a i_V (\theta)+i_V \theta',
\mbox{~~$\theta, \theta' $generic forms}\nonumber\\
& &v)~~~i_{\lambda V} = \lambda i_V,
\mbox{~~$\lambda \in \Cb$}\nonumber\\
& &vi)~~i_{\epsi t_i} (\theta)=\con{i}
(\theta) \epsi, \mbox{~~$\epsi \in A$}
\label{conprop}
\ena
As a consequence, the $q$-Lie derivative satisfies:
\eqa
& &i)~~~\ell_{\epsi t_i} a = i_{\epsi t_i} (a)=
\con{i} (a) \epsi \equiv (\epsi t_i) (a)
\nonumber\\
& &ii)~~\ell_V d \theta=d \ell_V \theta \nonumber\\
& &iii)~\ell_V (\lambda \theta +
\theta')=\lambda \ell_V (\theta)+ \ell_V (\theta')
\nonumber\\
& &iv)~~\ell_{\epsi t_i}(\theta)=(\lie{i}
\theta)\epsi - (-1)^p  \con{i}(\theta) d\epsi,
\mbox{~~$\theta$ generic p-form} \nonumber\\
& &v)~~~\ell_{ t_i}(\theta \we \om^k)=
(\lie{r} \theta) \we \om^l \L{rk}{li} +
\theta \we \lie{i} \om^k   \label{lieprop}
\ena
In analogy with the classical case, we define
the $q$-diffeomorphism variation of the fundamental field $\om^i$
as
\eq
\delta \om^k \equiv \ell_{\epsi^i t_i}  \om^k
\en
\noi where according to iv) in (\ref{lieprop}):
\eq
\ell_{\epsi^i t_i} \om^k= (\con{i}
d \om^k+d \con{i} \om^k) \epsi^i + d\epsi^k=
(\con{i} d\om^k )\epsi^i+ d \epsi^k
\en
As in the classical case, there is a
suggestive way to write this variation:
\eq
\ell_{\epsi^i t_i} \om^k= i_{\epsi^i t_i} R^k+ \nabla \epsi^k
\en
\noi where
\eqa
& &\nabla \epsi^k \equiv d\epsi^k -
\c{rs}{k} \con{i} (\om^r \we \om^s) \epsi^i=
\nonumber\\
& &~~~~~~de^k-\c{rs}{k} \epsi^r \om^s+\c{rs}{k} \om^r \epsi^s
\ena
\noi Proof: use the Bianchi
identities (\ref{bianchi}) and iii) in (\ref{conprop}).
\sk
\noi Notice that if we postulate:
\eqa
& &\L{rk}{li} \om^l \epsi^i = \epsi^r \om^k \nonumber\\
& &\L{rk}{li} \om^l \we d\epsi^i = -d\epsi^r \we \om^k \label{epsiom}
\ena
\noi we find
\eq
\de (\om^j \we \om^k)=\de\om^j \we
\om^k + \om^j  \we \de \om^k \label{omvar}
\en
\noi i.e. a rule that any ``sensible"
variation law should satisfy. To prove
(\ref{omvar}) use iv) and v) of (\ref{lieprop}). The $q$-commutations
(\ref{epsiom}) were already proposed in \cite{Cas1} in the context
of $q$-gauge theories.
\sk
We have now all the tools we need to investigate the invariances of the
$q$-gravity lagrangian (\ref{lagrangian}). These will be discussed in
a forthcoming publication. We anticipate
the result, analogous to the classical one:
after imposing the zero torsion constraint
$R^a=0$ and the horizontality conditions $\con{ab} R^{cd}=\con{ab} R^c=0$
along the Lorentz directions one finds that, {\sl provided}
the $\epsi$ tensor in  (\ref{lagrangian}) is appropriately defined,
the lagrangian is invariant under $q$-diffeomorphisms
and local Lorentz rotations. The correct definition of the
$q$-alternating tensor is:
\eq
\begin{array}{llll}
\epsi_{1234}=1,&\epsi_{1243}=-q,
&\epsi_{1324}=-1,&\epsi_{1342}=q,
\\
\epsi_{1423}=q^{3\over 2},&\epsi_{1432}=
-q^{3\over 2},&\epsi_{2134}=-1,&\epsi_{2143}=q,
\\
\epsi_{3124}=1,&\epsi_{3142}=-q,
&\epsi_{4123}=-q^{3\over 2},
&\epsi_{4132}=q^{3\over 2},
\\
\epsi_{2314}=q^{1\over 2},&\epsi_{2341}=
-q^{5 \over 2},&\epsi_{2413}=-q^2,&\epsi_{2431}=q^3,
\\
\epsi_{3214}=-q^{1\over 2},&\epsi_{3241}=
q^{5\over 2},&\epsi_{4213}=q^2,&\epsi_{4231}=-q^3,
\\
\epsi_{3412}=q^2,&\epsi_{3421}=-q^3,
&\epsi_{4312}=-q^2,&\epsi_{4321}=q^3,
\end{array}
\en
\sk
{\sl Note 1:} using the general formula (\ref{omb}) one sees that
the rule (\ref{epsiom}) is equivalent to postulating the following
coproduct on $\epsi^i$:
\eq
\D (\epsi^i)=\epsi^j \otimes \M{j}{i}
\en

{\sl Note 2:} The $q$-Lie derivative was
defined along left-invariant vectors
in ref. \cite{Aschieri1}, and extended to a Lie derivative along any
 tangent vector in ref. \cite{Aschieri2}. In
both these references, formulas
 are given where $t_i$ and $\om^j$ are
left-invariant. In (\ref{conprop}) and
(\ref{lieprop}) we have generalized these formulas to non-left invariant
$t_i$ and $\om^j$, with the $t_i$ still dual to the $\om^j$.
\sk
{\sl Note 3:} The $D=2$ bicovariant  $q$-Poincar\'e algebra proposed
in  the first of ref.s \cite{Cas} coincides with the one obtained from
 $SO_{q,r=1}(4)$ via
the procedure of Section 3.
\sk
{\bf Acknowledgements}
\sk
\noi It is a pleasure to acknowledge
useful discussions with Paolo Aschieri
and Branislav Jur\v{c}o.

\vfill\eject

\app{Change of basis and graphical representation of $\Lambda$}

Consider (\ref{qLie}) with adjoint indices:
\eq
\chi_i \chi_j-\L{kl}{ij} \chi_k \chi_l = \C{ij}{k} \chi_k \label{qLiead}
\en
Under a (nonsingular) change of basis
\eq
\chi_i = \Smat{i}{j} \xi_{j} \label{newchibasis}
\en
\noi the $q$-Lie algebra transforms into:
\eq
\xi_i \xi_j-\Lt{kl}{ij} \xi_k
\xi_l = \Ct{ij}{k} \xi_k \label{qLieadtrans}
\en
\noi with
\eq
\Lt{kl}{ij}= \Smatinv{i}{\ip} \Smatinv{j}{\jp} \L{\kp\lp}{\ip\jp}
\Smat{\kp}{k} \Smat{\lp}{l}
\en
\eq
\Ct{ij}{k}= \Smatinv{i}{\ip} \Smatinv{j}{\jp}\C{\ip\jp}{\kp}\Smat{\kp}{k}
\en
\noi which amounts to say that $\Lambda$ and $\Cb$ transform
as tensors under (\ref{newchibasis}).
Therefore we have the
\sk
{\sl Theorem:}
The transformed ${\tilde \Lambda}$ and $\tilde \Cb$ tensors
satisfy the bicovariance relations (\ref{bic1})-(\ref{bic4}).

{\sl Proof}: obvious since (\ref{bic1})-(\ref{bic4})
are tensor relations.
\sk
There is a particular change of basis for the $\chi$ that allows
a graphical representation for the braiding matrix $\Lambda$.
In the case of the $B,C,D$ series the new $\xi$ are defined via
the metric $C$:
\eq
\xi_{ab} = C_{ac} \cchi{c}{b}  \label{newchibasisab}
\en
\noi and the $\Lambda$ tensor takes the form:
\eq
\Lambda^{a_1 a_2 d_1 d_2}_{~~~~~~~~~~c_1 c_2 b_1 b_2}=
\Rhatinv{hg_1}{b_1c_2} \Rhatinv{e a_1}{g_1c_1} \Rhat{d_1a_2}{fe}
\Rhat{d_2f}{b_2h}  \label{Lambdaud}
\en
\noi To prove this, one uses the definition (\ref{Dmatrix})
inside (\ref{Lambda}), and the identities (\ref{crc1})-(\ref{crc2}).

If we represent
$\Rh$ and $\Rh^{-1}$ as

\begin{center}
\begin{picture}(300,50)
   \put(0,0){\begin{picture}(100,50)
              \put(0,0){$\Rhat{ab}{cd}~=$}
              \thicklines
              \put(60,-20){\begin{picture}(50,50)
                             \put(10,10){\line(1,1){12}}
                             \put(10,40){\line(1,-1){30}}
                             \put(40,40){\line(-1,-1){12}}
                             \put(0,0){c} \put(0,42){a}
                             \put(40,0){ d} \put(40,42){ b}
                           \end{picture}}
             \end{picture}}

   \put(150,0){\begin{picture}(100,50)
              \put(0,0){$\Rhatinv{ab}{cd}~=$}
              \thicklines
              \put(60,-20){\begin{picture}(50,50)
                             \put(10,10){\line(1,1){30}}
                             \put(10,40){\line(1,-1){12}}
                             \put(40,10){\line(-1,1){12}}
                             \put(0,0){c} \put(0,42){a}
                             \put(40,0){ d} \put(40,42){ b}
                           \end{picture}}
             \end{picture}}
\end{picture}
\end{center}

\vfill\eject

\noi then the braiding
matrix $\Lambda$ is represented by (cf. also \cite{Watamura}):

  \begin{picture}(300,150)
         \put(0,75)
           {$\Lambda^{a_1 a_2 d_1 d_2}_{~~~~~~~~~~c_1 c_2 b_1 b_2}~=$}
         \thicklines
         \put(150,25){\begin{picture}(150,100)
                             \put(20,20){\line(1,1){27}}
                             \put(53,53){\line(1,1){10}}
                             \put(80,80){\line(-1,-1){12}}

                             \put(50,20){\line(1,1){60}}

                             \put(20,80){\line(1,-1){42}}
                             \put(80,20){\line(-1,1){12}}

                             \put(50,80){\line(1,-1){27}}
                             \put(110,20){\line(-1,1){27}}
                         \put(10,10){$b_2$}
                         \put(40,10){$b_1$}
                         \put(80,10){$c_2$}
                         \put(110,10){$c_1$}
                         \put(10,85){$d_2$}
                         \put(40,85){$d_1$}
                         \put(80,85){$a_2$}
                         \put(110,85){$a_1$}
                    \end{picture}}
   \end{picture}

The metric $C_{ab}$ is represented as

\begin{center}
\begin{picture}(300,50)
   \put(0,0){\begin{picture}(100,50)
              \put(0,0){$C_{ab}~=$}
              \thicklines
              \put(60,-10){\begin{picture}(50,50)
                             \put(25,20){\oval(25,25)[t]}
                              \put(10,10){$a$}
                              \put(35,10){$b$}
                          \end{picture}}
             \end{picture}}

   \put(150,0){\begin{picture}(100,50)
              \put(0,0){$C^{ab}=$}
              \thicklines
              \put(60,-10){\begin{picture}(50,50)
                             \put(25,20){\oval(25,25)[b]}
                              \put(10,25){$a$}
                              \put(35,25){$b$}
                           \end{picture}}
             \end{picture}}
\end{picture}
\end{center}

\noi and, as an amusing exercise,  the reader can draw
the graphical representation of the adjoint metric $C_{ij}$ given
in (\ref{metricadjoint}) and its inverse, and of the
relations (\ref{crc1})-(\ref{crc2}) and (\ref{CRC1})-(\ref{CRC2}).
The metric $C_{ab}$ allows to close the braids into knots, and
the graphical representation of this Section yields a knot
invariant for any $B_n,C_n,D_n$ $q$-group.
 The three Reidemeister moves hold because of

i) $C_{ab} \Rhat{ab}{cd}
\propto C_{cd}$,

ii) the definition of the crossings corresponding
to $\Rh$ and $\Rh^{-1}$,

iii) the quantum Yang-Baxter equations
for $\Rh$.

\noi On the connection between knot theory and quantum groups see
ref.s \cite{knots,Watamura}.

\sect{A note on bicovariance conditions and $q$-groups defined in the
adjoint representation}

Whenever we have a set of $\Lambda$ and $\Cb$ components
satisfying the bicovariance conditions (\ref{bic1})-(\ref{bic4}),
and an invertible  metric $C_{ij}$ satisfying (\ref{CRC1}), (\ref{CRC2}),
we can define
\sk
i) a quantum group generated by the matrix elements $\M{i}{j}$.
The ``$RTT$" relations become now ``$\Lambda MM$" relations:

\eq
\M{i}{j} \M{r}{q} \L{ir}{pk}=\L{jq}{ri} \M{p}{r} \M{k}{i} \label{LMM}
\en

The co-structures on $M$ are similar to the ones defined on the $T$
(eqs. (\ref{cos1})-(\ref{cos2})):

\eqa
& & \D(\M{i}{j})=\M{i}{k} \otimes \M{k}{j} \\
& & \epsi (\M{i}{j})=\delta^j_i\\
& & \kappa(\M{i}{j})= C_{ik} \M{l}{k} C_{lj} \label{Mantipode}
\ena
Moreover, we can impose the orthogonality relations:
\eq
\M{i}{j} \M{k}{l} C^{ik} = C^{jl}
\en
\eq
\M{i}{j} \M{k}{l} C_{jl}=C_{ik}
\en
\noi These are compatible with the $\Lambda M M$
relations (\ref{LMM}) because of eq.s (\ref{CRC1}) and (\ref{CRC2}).
\sk

ii) functionals $\chi_i$ and $\f{i}{j}$ via their action on $M$:

\eq
\chi_j (\M{i}{k})=\C{ij}{k}
\en
\eq
\f{i}{l} (\M{k}{j})=\L{ij}{kl}
\en
\noi whose co-structures are given by:
\eqa
& & \Dp (\chi_i)=\chi_j \otimes \f{j}{i} + \Ip \otimes \chi_i \\
& & \ep (\chi_i)=0 \\
& & \kappa^{\prime}
(\chi_i)= - \chi_j \kappa^{\prime} (\f{j}{i}) \label{chiantipode}
\ena
\eqa
& & \Dp (\f{i}{j})=\f{i}{k} \otimes \f{k}{j} \\
& & \ep(\f{i}{j})=\de^i_j \\
& & \kappa^{\prime} (\f{i}{j})= \f{i}{j} \circ \kappa
\ena

The action of $\chi_i$ and $\f{i}{j}$ on products of $M$ elements
is defined in the usual way via the coproduct $\Dp$, i.e.
$\chi_i (ab)=\Dp (\chi_i) (a \otimes b)$ etc.

These functionals satisfy the relations:

\eq
\chi_i\chi_j-\L{kl}{ij} \chi_k\chi_l=\C{ij}{k} \label{bicop1}
\en
\eq
\L{nm}{ij} \f{i}{p} \f{j}{q}=\f{n}{i} \f{m}{j} \L{ij}{pq} \label{bicop2}
\en
\eq
\C{nm}{i} \f{m}{j} \f{n}{k}+\f{i}{j} \chi_k=\L{pq}{jk} \chi_p
\f{i}{q}+\C{jk}{l} \f{i}{l}
\en
\eq
\chi_k \f{n}{l} = \L{ij}{kl} \f{n}{i} \chi_j \label{bicop4}
\en

\noi which are the operatorial equivalents of the bicovariance
relations (\ref{bic1})-(\ref{bic4}), cf. \cite{Bernard}.
Indeed the latter can be
obtained by applying the former to $\M{r}{s}$. We recall that
products of functionals are convolution products, for example
\eq
\chi_i \chi_j \equiv (\chi_i \otimes \chi_j) \D
\en

The algebra generated by the $\chi$ and $f$ modulo the relations
(\ref{bicop1})-(\ref{bicop4}) is a Hopf algebra, and defines
a bicovariant differential calculus on the $q$-group generated by
the $M$ elements. For example, one can introduce left-invariant
one-forms $\om^i$ as duals to the  ``tangent vectors" $\chi_i$,
an exterior product
\eq
 \om^i \we \om^j \equiv \om^i \otimes \om^j -
\L{ij}{kl} \om^k \otimes \om^l ,\label{wedge}
\en
\noi an exterior derivative
on $Fun_q(\M{i}{j})$ (the quantum group generated by
the $\M{i}{j}$) as
\eq
da= (id \otimes \chi_i) \D (a) \om^i, \mbox{~~~$a \in Fun_q(\M{i}{j})$  }
\en
\noi and so on.  The commutations between one-forms and elements
 $a \in Fun_q(\M{i}{j})$ are given by:
\eq
\om^i a=(id \otimes \f{i}{j}) \D (a) \label{omb}
\en
\sk
{\sl Note:}  when $\Lambda^2=1$ there is a way to define the antipode
of $\f{i}{j}$ without reference to an adjoint metric. This we manage by
enlarging the algebra, adding  to
$\chi_i$ and $\f{i}{j}$ also the functionals $\fm{i}{j}$
defined by:
\eq
\fm{i}{l} (\M{k}{j})=\L{ji}{lk}
\en
\noi An antipode for $\f{i}{j}$ can be defined as
\eq
\kappa^{\prime} (\f{i}{j})=\fm{i}{j} \label{fantipode}
\en
\noi since
\eq
\fm{i}{s} \f{s}{j} (\M{l}{k})=(\fm{i}{s} \otimes \f{s}{j}) \D (\M{l}{k})=
\fm{i}{s} (\M{l}{r}) \f{s}{j}
(\M{r}{k})=\L{ri}{sl} \L{sk}{rj}=\de^i_j \de^k_l
\en
\noi the last equality being due to eq. (\ref {Ltilde}). Then we see that
\eq
\fm{i}{s} \f{s}{j} =\de^i_j \epsi
\en
\noi so that $\fm{i}{j}$ is a good
inverse for $\f{i}{j}$. Similarly we can prove
that
\eq
\f{i}{s} \fm{s}{j} (\M{l}{k})=\de^i_j \de^k_l
\en
\noi and therefore
\eq
\kappa^{\prime} (\fm{i}{j})=\f{i}{j}
\en
\noi which means that i) $\kappa'^2=1$ , ii) the algebra generated
by $\chi_i, \f{i}{j}, \fm{i}{j}$
is closed under all the co-structures (actually
the set of the generators itself is closed).
\sk
Using eq. (\ref{fantipode}),  the relations (\ref{bicop1})-(\ref{bicop4})
can be easily extended  to include the
$\fm{i}{j}$.

\sect{Extension of the $*$-conjugation to the dual algebra
generated by the $\chi$ and $f$ functionals}

Here we want to show that the conjugation:
\eq
(\chi_i)^* = - \Dc_{i}^{~\ip} \chi_{\ip} \label{chistar}
\en
\eq
(\f{i}{j})^*= \Dc_{j}^{~\jp} \f{\ip}{\jp} \Dc_{\ip}^{~i} \label{fstar}
\en
\noi where $\Dc$ is the matrix that permutes the
fundamental indices $n \leftrightarrow n+1$,
is compatible with the operatorial bicovariance
conditions (\ref{bicop1})-(\ref{bicop4}). We begin
by taking the $*$-conjugate of (\ref{bicop1}), which yields:
\eq
\chi_j^* \chi_i^* - \Lbar{kl}{ij} \chi_l^* \chi_k^*=
\Cbar{ij}{k} \chi_k^*
\en
\noi Using now (\ref{chistar}) and the fact that ${\bar \Lambda}$
is invariant under the index permutation $n \leftrightarrow n+1$
we can rewrite the above equation as:
\eq
\chi_j \chi_i - \Lbar{kl}{ij} \chi_l\chi_k=
-\Cbar{ij}{m} \chi_m \label{qlieconj}
\en
{\sl Remark:} for $|q|=1$ we have
\eq
\Lbar{kl}{ij}=\L{lk}{ji} \label{c1}
\en
Indeed rewrite  (\ref{bicop1}) as
\eq
\chi_j \chi_i - \L{lk}{ji} \chi_l \chi_k =
\C{ji}{m}  \chi_m\label{qliereversed}
\en
\noi For $r=1$, $\L{lk}{ji}$ is proportional to $\de^k_j \de^l_i$
(and contains only one term, as we can see from the defining
formula  (\ref{Lambda}) and the fact that $R$ is diagonal with
only one term in the diagonal entries)
and it is easy to deduce that (\ref{qliereversed}) is compatible with
(\ref{bicop1}) only if $\L{lk}{ji}=(\L{kl}{ij})^{-1}$ (the inverse of
the matrix {\it element}). For $|q|=1$
we have $\Lbar{kl}{ij}=(\L{kl}{ij})^{-1}$
so that  (\ref{c1}) is proved.
\sk
Notice now  that (\ref{qlieconj})
would reproduce exactly (\ref{qliereversed}),
proving the compatibility of the
conjugation rule (\ref{chistar}) with the
$q$-Lie algebra , if we had also:
\eq
\Cbar{ij}{m}=-\C{ji}{m} \label{c2}
\en
We could not prove (\ref{c2}) on general grounds: presumably one can
always find a basis for the $\chi_i$ generators such that
it holds. This is indeed  the case for $ISO_q(3,1)$.
\sk
In a similar way one shows the compatibility of the
conjugation (\ref{chistar}), (\ref{fstar}) with the remaining
bicovariance operator conditions (\ref{bicop2})-(\ref{bicop4}).
One just needs to use property (\ref{c2}) and a useful identity valid for
$\Lambda^2=I$:
\eq
\C{ij}{m} = - \L{kl}{ij} \C{kl}{m} \label{c3}
\en
\noi i.e. the structure constants are $\Lambda$-antisymmetric.
This is easily proved by remarking that the left-hand
side of (\ref{bicop1}) does not change under multiplication
by the projector $P_A=(I-\Lambda)/2$, since $P_A (I-\Lambda)=I-\Lambda$
(for $\Lambda^2=I$), and thus
\eq
 {1 \over 2} (I-\Lambda)^{kl}_{~~ij} \C{kl}{m}=\C{ij}{m}
\en
\noi which is just eq. (\ref{c3}). Again this can be immediately checked
to hold for  $ISO_q(3,1)$.
\sk
Finally, it is a simple matter to check that
\eq
\kappa ' (\kappa ' (\chi_i^*)^*)=\chi_i,~~~\Dp(\chi_i^*)=[\Dp(\chi_i)]^*
\en
\noi (and similar for $\f{i}{j},\fm{i}{j}$)
showing that (\ref{chistar})-(\ref{fstar})
is a Hopf algebra conjugation.

\vfill\eject

\centerline{\bf Table 1 }
\centerline{ $ISO_{q}(N) $ Lie algebra}
\eqa
& &\cchi{c_1}{c_2} \cchi{b_1}{b_2} -
\LL{a_1}{a_2}{d_1}{d_2}{c_1}{c_2}{b_1}{b_2}~
\cchi{a_1}{a_2}\cchi{d_1}{d_2}=\CC{c_1}{c_2}{b_1}{b_2}{d_1}{d_2}~
\cchi{d_1}{d_2} ~~~\mbox{[$SO_{q,r=1}(N)$ Lie algebra]} \nonumber \\
& &\cchi{c_1}{c_2} \cchi{b_1}{\circ} -
 {q_{\circ d_1 }\over q_{\circ d_2}}
\R{d_2b_1}{c_2g_1} \Rinv{c_1g_1}{d_1a_1}~
\cchi{a_1}{\circ}\cchi{d_1}{d_2}=\R{g_2b_1}{c_2g_1}
\Rinv{c_1g_1}{e_1a}\Rinv{ae_1}{g_2d_1} ~
\cchi{d_1}{\circ} \nonumber \\
& &\cchi{c_1}{\circ} \cchi{b_1}{\circ} -
   {q_{\circ d_1 }\over q_{\circ b_1}}\Rinv{c_1b_1}{d_1a_1} ~
\cchi{a_1}{\circ}\cchi{d_1}{\circ}=0 \nonumber \\
& &\cchi{c_1}{c_2} \cchi{\bullet}{b_2} -
{q_{\circ c_1 }\over q_{\circ c_2}}
\Rinv{a_2c_1}{g_2d_1} \R{g_2d_2}{b_2c_2}~
\cchi{\bullet}{a_2}\cchi{d_1}{d_2}= {q_{\circ c_1} \over q_{\circ c_2}}
\R{c_1d_2}{b_2c_2} ~
\cchi{\bullet}{d_2} \nonumber \\
& &\cchi{\bullet}{c_2} \cchi{\bullet}{b_2} -
{q_{\circ a_2 }\over q_{\circ c_2}}\R{a_2d_2}{b_2c_2} ~
\cchi{\bullet}{a_2}\cchi{\bullet}{d_2}=0 \nonumber \\
& &\cchi{c_1}{\circ} \cchi{\bullet}{b_2} -
q_{\circ c_1 } q_{\circ b_2} \Rinv{a_2c_1}{b_2d_1}  ~
\cchi{\bullet}{a_2}\cchi{d_1}{\circ}=0 \nonumber
\ena
\sk\sk
\centerline{\bf Table 2}
\centerline{ $ISO_{q}(N) $ Lie algebra in the
$\chi^a \equiv q_{\circ a'}^{1\over 2} \cchi{a}{\circ}-
q_{\circ a'}^{-{1\over 2}}
\cchi{\bullet}{a'}$
basis}
\eqa
& &\cchi{c_1}{c_2} \cchi{b_1}{b_2} -
\LL{a_1}{a_2}{d_1}{d_2}{c_1}{c_2}{b_1}{b_2}~
\cchi{a_1}{a_2}\cchi{d_1}{d_2}=\CC{c_1}{c_2}{b_1}{b_2}{d_1}{d_2}~
\cchi{d_1}{d_2} ~~~\mbox{[$SO_{q,r=1}(N)$ Lie algebra]} \nonumber \\
& &\cchi{c_1}{c_2} \chi^{b_1} -
 {q_{\circ d_1 }\over q_{\circ d_2}}
\R{d_2b_1}{c_2g_1} \Rinv{c_1g_1}{d_1a_1}~
\chi^{a_1}\cchi{d_1}{d_2}=
{q^{1\over 2}_{\circ b_1'}\over {q^{1\over 2}_{\circ d_1'} } }
\R{g_2b_1}{c_2g_1}
\Rinv{c_1g_1}{e_1a}\Rinv{ae_1}{g_2d_1} ~
\chi^{d_1} \nonumber \\
& &\chi^{c_1}\chi^{b_1} -
   {q_{\circ d_1 }\over q_{\circ b_1}}\Rinv{c_1b_1}{d_1a_1} ~
\chi^{a_1}\chi^{d_1}=0 \nonumber
\ena
\centerline{\bf Table 3}
\centerline{$ISO_q(3,1)$ Lie algebra
in the $\chi_a \equiv C_{ab} \chi^b$,
$\chi_{ab}\equiv {1\over 2} [C_{ac} \cchi{c}{b}-C_{bc}
\cchi{c}{a}]$ basis}
\eqa
& &[\chi_{ab},\chi_{cd}]=C_{bc} \chi_{ad} + C_{ad} \chi_{bc}-
         C_{bd} \chi_{ac}-C_{ac} \chi_{bd} \nonumber \\
& &[\chi_{12}, \chi_a]_{q^{-1}} =\qmh C_{2a} \chi_1-\qmh C_{1a} \chi_2
       \nonumber \\
& &[\chi_{13}, \chi_a]_{q^{-1}} =\qmh C_{3a} \chi_1-\qmh C_{1a} \chi_3
       \nonumber \\
& &[\chi_{14}, \chi_a] =C_{4a} \chi_1-C_{1a} \chi_4
       \nonumber \\
& &[\chi_{23}, \chi_a] =C_{3a} \chi_2-C_{2a} \chi_3
       \nonumber \\
& &[\chi_{24}, \chi_a]_q =\qh C_{4a} \chi_2-\qh C_{2a} \chi_4
       \nonumber \\
& &[\chi_{34}, \chi_a]_q =\qh C_{4a} \chi_3-\qh C_{3a} \chi_4
       \nonumber \\
& &[\chi_1,\chi_2]_{\qm}=0,~~~~~~~[\chi_1,\chi_3]_{\qm}=0 \nonumber\\
& &[\chi_1,\chi_4]_{\qmt}=0,~~~~~~~[\chi_2,\chi_3]=0 \nonumber\\
& &[\chi_2,\chi_4]_{\qm}=0,~~~~~~~[\chi_3,\chi_4]_{\qm}=0 \nonumber
\ena
{}~~~~~with $[A,B]_s \equiv AB-sBA$.
\vfill\eject

\vfill\eject
\end{document}